\documentstyle[aps,preprint]{revtex}
\input epsf

\title{Vortex shedding in a model of superflow}

\author{C. Josserand\thanks{Present Adress: The James Franck Institute,
The University of Chicago, 5640 South Ellis Av., Chicago, Illinois 60637, USA.},
Y. Pomeau
\& S. Rica\\
\small\it LPS, laboratoire associ\'e au CNRS, Ecole Normale Sup\'erieure,
24, rue Lhomond, 75231 Paris Cedex 05, France\\ \small \it
and Laboratoire ASCI UPR 9029, Orsay, France.
}

\author{\parbox{430pt}{\vglue 0.3 cm \small    
The present article
represents part of the PhD. dissertation by C. Josserand \cite{these}.  
We discuss the nucleation of quantized vortices in the nonlinear 
Schr\"{o}dinger equation (NLS) for a flow around a disk in
two spatial dimensions. It appears that the vortices are nucleated when the flow 
becomes locally (at the edge of the disk) supersonic. A detailed study of the
phase equation for the complex field $\psi$ gives an Euler--Tricomi type equation
for the stationary solutions below threshold. This equation is closely related 
to the one known in shock wave dynamics for gas. Then using solvability 
condition, we extract a time--dependent scenario for the evolution of the
amplitude of the solution, which we, finally, relate to a known family solution
of NLS which  gives rise to a vortex nucleation. We also give a first order
correction  at the Landau velocity of nucleation, taking into account the
geometry of the flow.
}}

\author{\parbox{430pt}{\vglue 0.3 cm \small PACS numbers: 74.20.De;
71.10.-w.}}

\begin{document}

\maketitle
\tableofcontents

\section{Introduction}

\begin{figure}[htp]
\centerline{a) \epsfxsize=7truecm \epsfbox{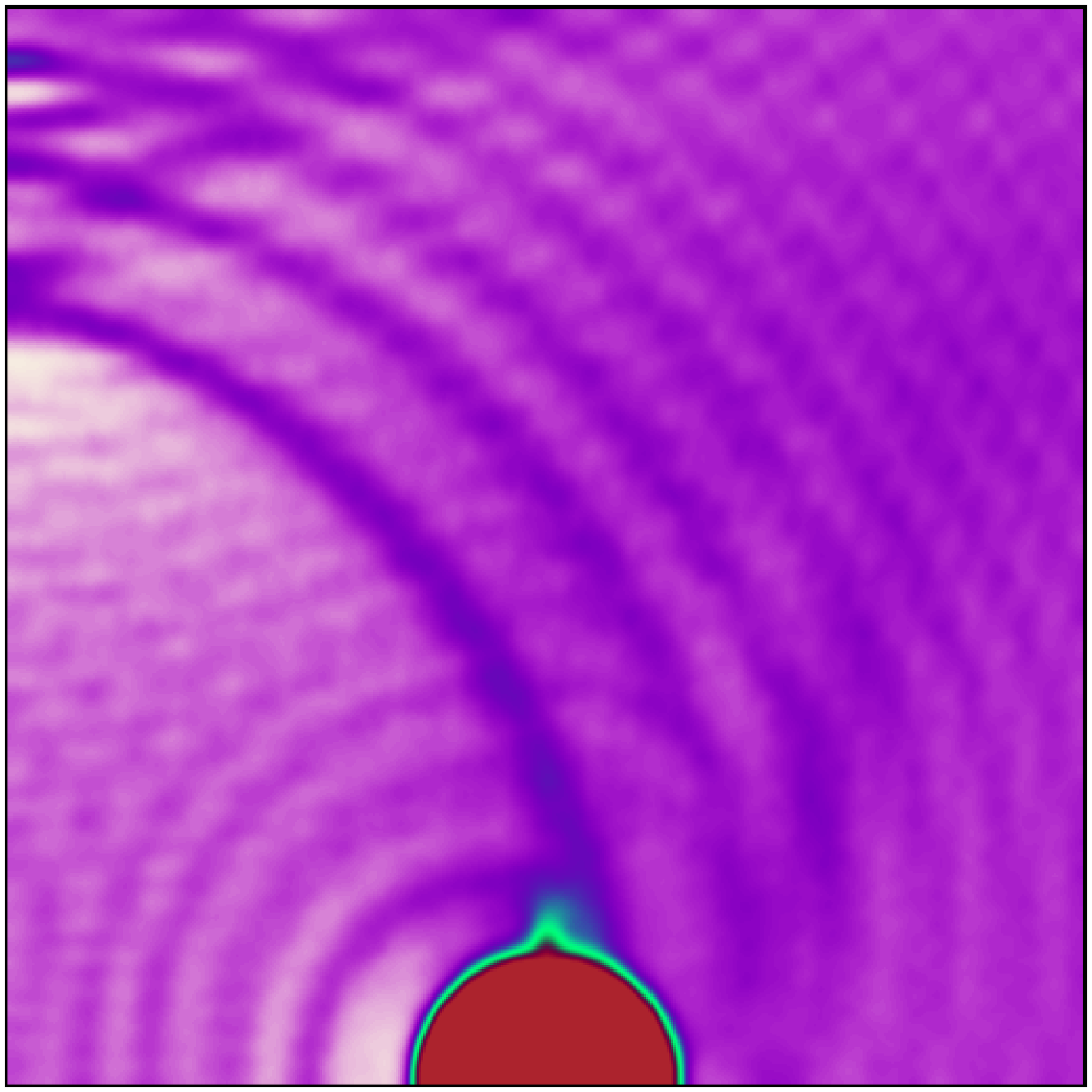} b) \epsfxsize=7truecm \epsfbox{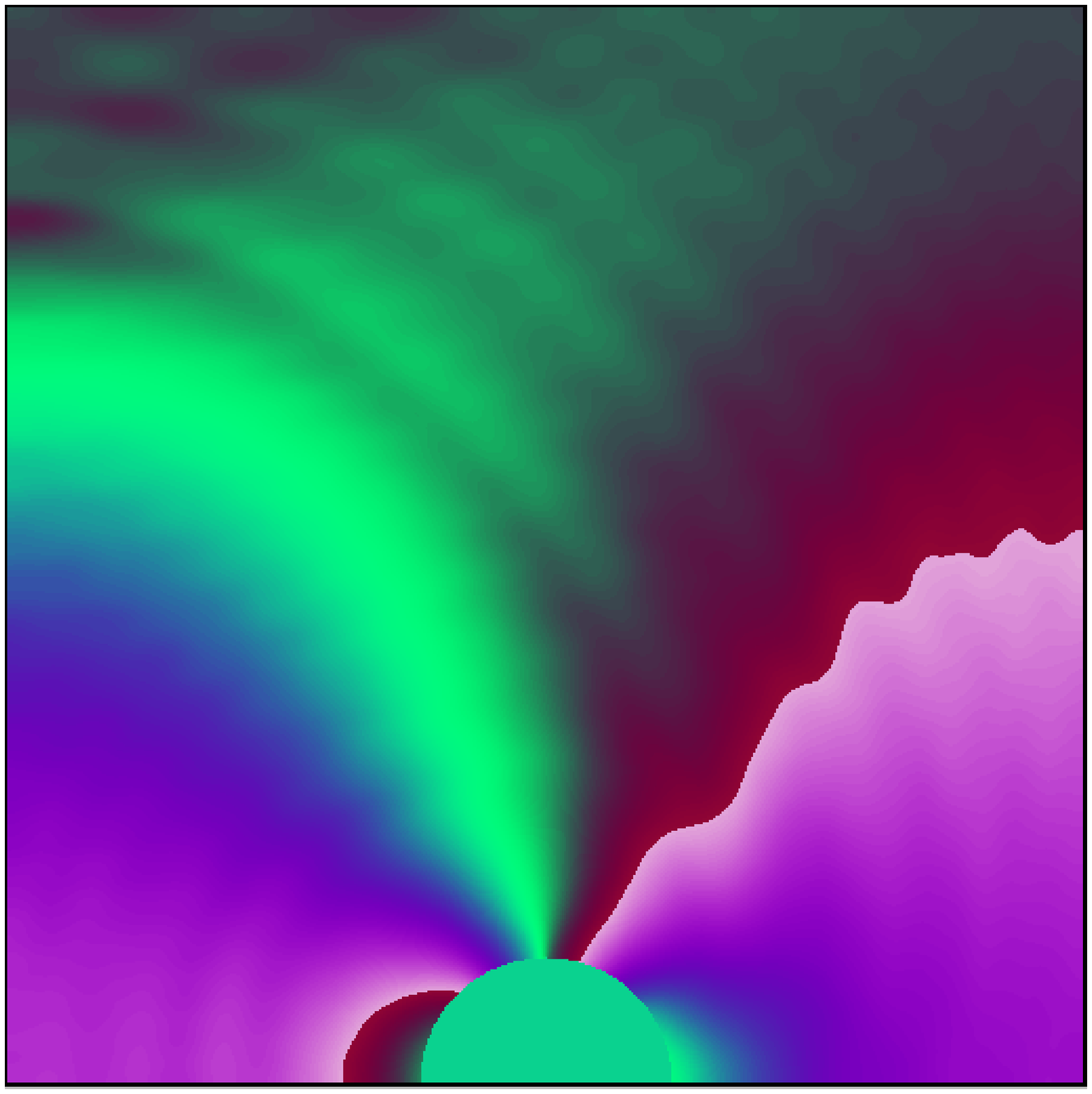} }
\centerline{c) \epsfxsize=7truecm \epsfbox{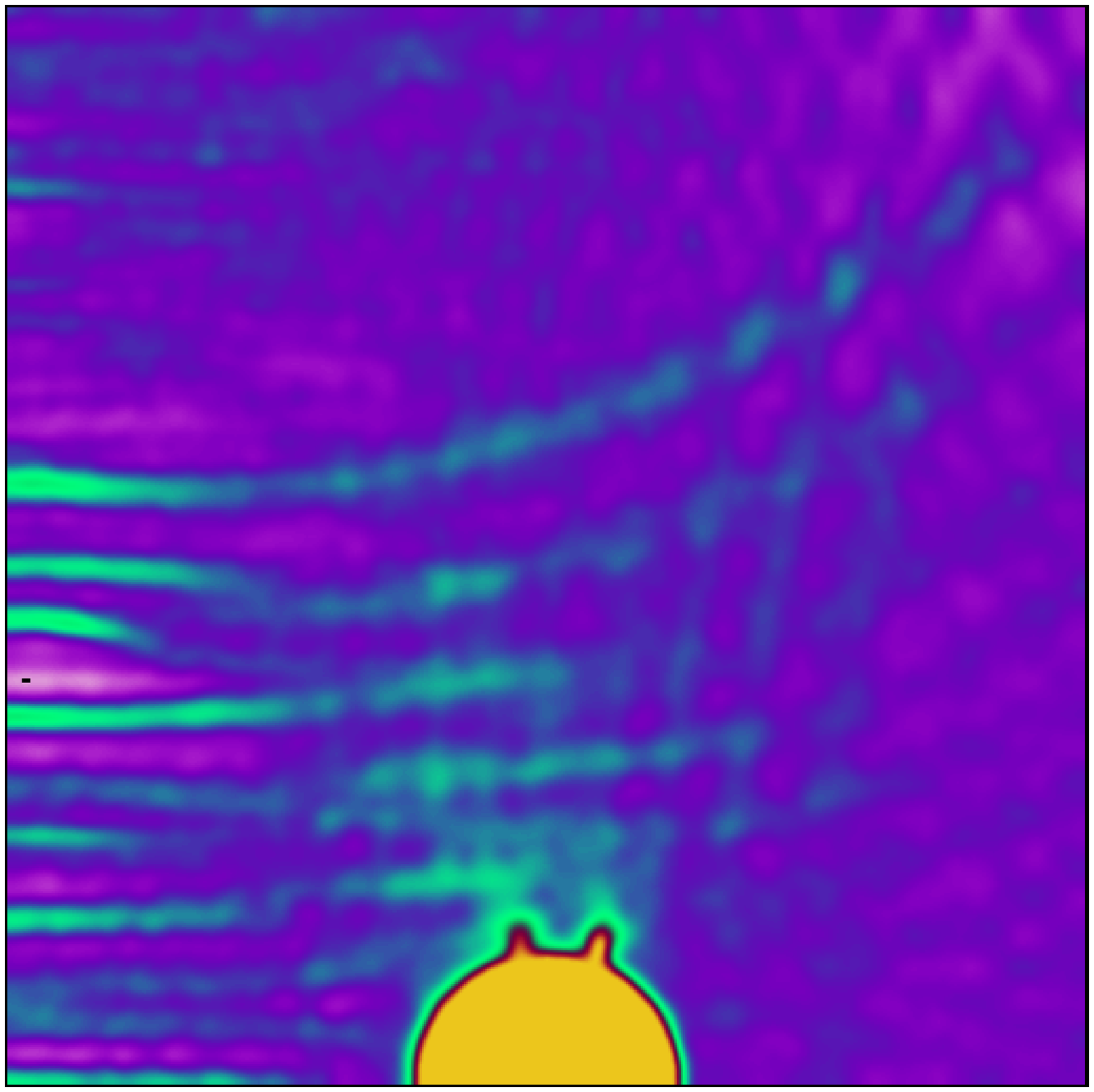} d) \epsfxsize=7truecm \epsfbox{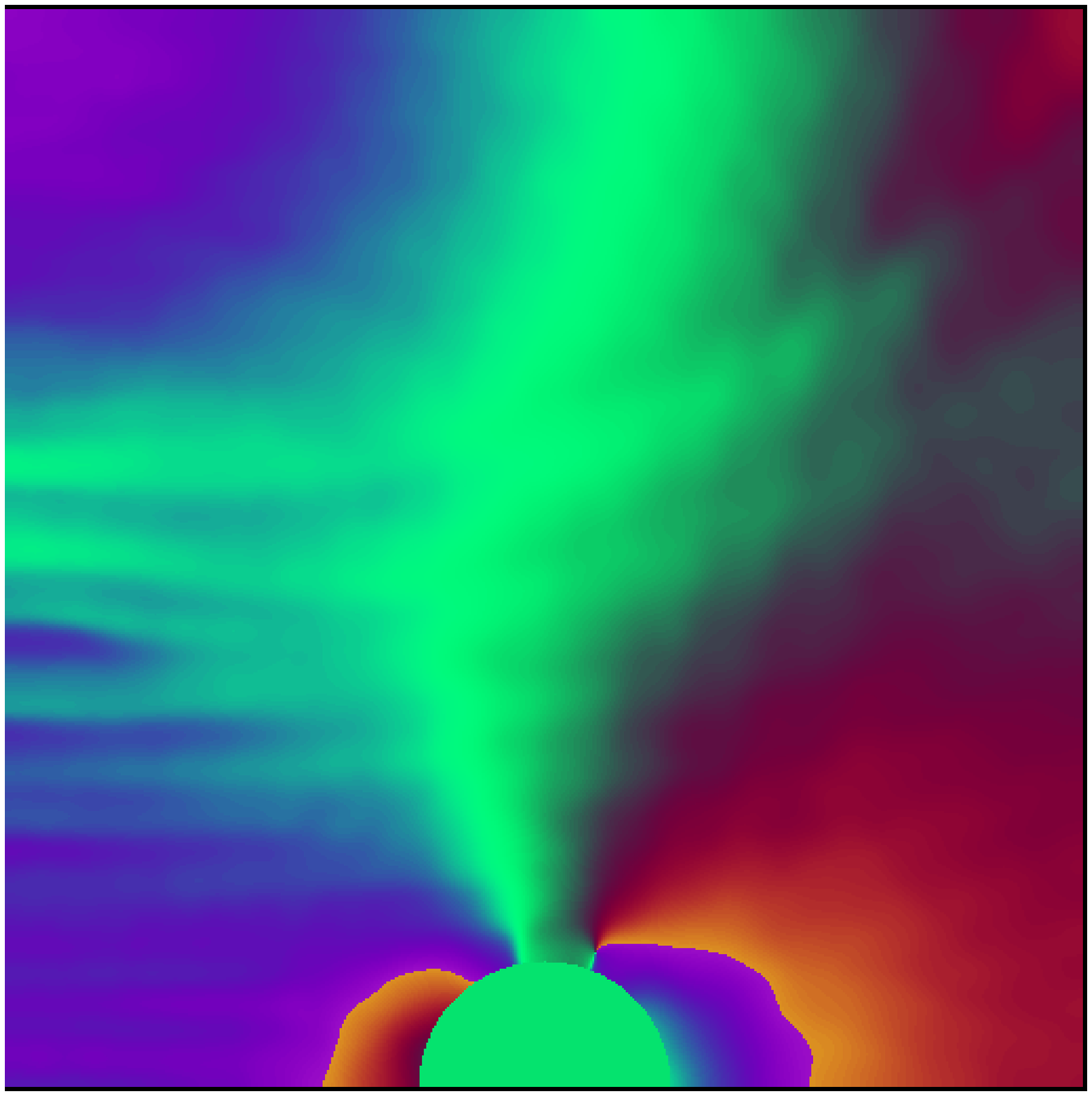}}
\caption{\protect\small Numerical simulation of the nonlinear
Schr\"{o}dinger equation for a bidimensional flow around half a disk;
the velocity at infinity is $v_\infty=0.442$ and we have taken $dx=0.125$ the
mesh grid and the radius of the disk is $R=7.5$.
a) \& b) respectively the modulus and the phase of the wave function at $t=20$
time unit of NLS. The densite and the phase go up from bright to dark color.
One can see the low density around the top of the obstacle, due to a
Bernoulli effect.
c) \& d) same functions at $t=50.6$. A low density structure is advected by the
flow (at right of the top of the disk). One can see a phase discontinuity 
and the tip of it where the phase is
not defined this is the signature of a topological defect, that is a
quantized vortex.
\label{photo}}
\end{figure}

The dynamics of superfluids at zero temperature can be modelled by the Gross-Pitaevski\v{\i} 
equation \cite{pit}. This is a partial differential equation for a complex valued scalar field
$\psi({\bf x},t)$ with dependence on the position in space, ${\bf x}$ and on time $t$. 
At zero temperature, there is no formal damping term in the equation; it is
reversible in time (after complex conjugation) and has even a Lagrangian structure. 
Although many of properties are well--known and have been studied quite 
extensively, we shall first review some of them. This
Gross-Pitaevski\v{\i} equation can be seen as a nonlinear Schr\"odinger equation, and 
so shares many properties of the linear Schr\"odinger equation. With 
periodic boundary conditions, it has an uniform ground state minimizing the 
energy, the corresponding solution
$\psi({\bf x},t)$ depends periodically on time through a simple exponential factor. 
As the Gross-Pitaevski\v{\i} equation is Galilean invariant, it is easy to 
construct a solution representing an uniform flow 
by boosting the rest state to a specified speed. With the same model, 
it is also possible to look at more complicated situations, 
as flows around obstacles. Two of the authors  \cite{fpr,cras} have studied over
the last years  a 2D flow around a circular disk (among others), that is the 
solution of the Gross--Pitaevski\v{\i} equation with a given uniform and constant
flow speed and mass density
at infinity, with a boundary condition on the surface of the disk. One 
striking observation has been that beyond a certain critical 
speed, the flow around the disk becomes time dependent, because vortices 
are emitted from the disk surface as shown on figure (\ref{photo}).

In Ref. \cite{fpr} it was shown that the release
of vortices from the boundary of the disk is a consequence of a transition from 
a locally subsonic to supersonic flow. In ordinary fluid mechanics, this would 
lead to the formation of a shock wave inside the so--called supersonic bubble.
Nothing resembling this is possible in the Gross--Pitaevski\v{\i} equation,
because of the lack of built--in irreversibility, something that is necessary to
balance nonlinearities inside the shock wave. As argued in Ref. \cite{fpr}, in
the present model, the formation of shock waves is replaced --albeit in a rather
loose sense-- by the nucleation of vortices with a quantized circulation. 
In fact, these
vortices are topological defects solutions of the Gross--Pitaevski\v{\i}
equation. In two space dimensions they are points where the complex field
$\psi({\bf x},t)$ vanishes, making a $\pm 2\pi$ phase jump when one turns around
the defect. These vortices are emitted  when the local velocity becomes larger
than the  critical velocity making the flow locally supersonic\footnote{ The
present problem  has been studied in the one dimension case by Hakim
\cite{vincent} who has shown that the release of vortices is replaced by a
periodic nucleation of one dimensional solitons.}. 

The present work is
devoted to study this transition. This is a rather intricate question, as anyone
who looked at  the paragraph \S118, \S119 \& \S120 of Landau and Lifschitz book
on fluid mechanics
\cite{landau} knows well: one has to solve locally the Euler--Tricomi equation for the
velocity  potential, that can be done in terms of the Airy 
function. In the present case there is a further complication with respect 
to Landau's analysis, which comes from the boundary conditions imposed on
the disk, see \S126 of Ref. \cite{landau}. Let us outline the
principles of our analysis; we are dealing with a problem, with two
small parameters: the ratio of the microscopic intrinsic length to the disk radius, and
the relative  shift of the velocity near the pole of the disk to the critical value, a
parameter called $\epsilon$.\footnote{$\epsilon$ is also proportional to the difference
between the actual Mach number and its critical value.}  Far from the disk boundary, in
terms of the microscopic distance, the usual  hydrodynamic assumption holds: the quantum
pressure term can be  neglected, and one obtains an unique, nonlinear equation for the
velocity  potential\footnote{In this hydrodynamical limit, the microscopic length scale
does not appear, and there is only one small parameter, $\epsilon$.}. This equation can be
solved by expansion, by assuming that  at the dominant order the surface of the disk is
flat. In this limit, the  hydrodynamic boundary conditions are satisfied with an uniform 
tangent velocity field. Corrections have to  be added to this velocity field in order to
take into  account the curvature of the disk boundary. The first correction is trivial, 
and only the second one is crucial for the analysis. 

The second order corrections (for the velocity potential) is the solution of the
Euler--Tricomi  equation, with specific boundary condition. It happens that the 
small parameter $\epsilon$ enters into this Euler--Tricomi equation plus the 
boundary conditions problem in such a way that it can be factored out by
rescalings. Moreover, for $\epsilon$ negative, that is for velocities slightly
less  than the critical speed, the velocity potential at this approximation 
is multivalued, but only in its extension inside the disk, which is a
non--physical part of the flow.  At  $\epsilon$ =0 there is a transition, and
the boundary of the  region where the velocity potential becomes multivalued
enters  the physical space, so that the solution of Euler--Tricomi equation
cannot  be considered as physically acceptable in this range of parameters. In 
ordinary viscous fluids, this multivaluedness would signal the formation 
of a shock wave. As said before, no such a thing exists in the 
present model of superfluid. Accordingly, the shock wave is replaced by 
a vortex, that is by adding to the flow field in the slightly supersonic 
region the flow field coming from a localized vortex close to the boundary. 
Taking in account all the nonlinear terms and the time dependent dynamics 
on the phase equation, we show that this lack of stationarity appears as
the result of a saddle--node bifurcation, as suggested by two of us some ago
\cite{cras}. We argue also that the critical  velocity behaves as
$\sqrt{\frac{\xi_0}{R}}$ above the transonic region, being $\xi_0$ the healing
length of NLS and $R$ the radius of  the disk; this is because the
quantum pressure stabilizes the flow  above the transonic transition for such a
range of velocity. We claim that this saddle--node bifurcation, giving rise to
time--dependent  dynamic  corresponds to vortex emission; for that, it remains 
to match this outer velocity potential, solution 
of the Euler--Tricomi equation with an inner solution, close to the disk boundary
with a vortex.

As one might expect from this rather lengthy introduction, it is a rather
uneasy job to put together all this picture, particularly because it depends 
on properties of solutions the Euler--Tricomi equation that are not obvious.
In section \ref{sec2}, we shall present the general problem in its explicit form,
that is the form of the equations as well as the boundary conditions and the
relevant facts about the hydrodynamic limit. We finally (section
\ref{sec2.5}) obtain a nonlinear equation  for an additional phase which can be
decomposed as the usual Euler--Tricomi  equation plus nonlinear terms that will
be treated as  perturbations.
In section \ref{sec3}, we study some properties of the Euler--Tricomi 
equation, particularly we point out the multivaluedness solutions of such 
an equation.
In section \ref{sec4}, we solve the Euler--Tricomi equation (including the
quantum  pressure term) using a Fourier transform 
along the $x-$axis first and finding the solution in term of the Airy function.
This gives a first correction to the critical velocity coming from the 
interplay between supersonic flow and regularisation due to the quantum 
pressure. Then, in section \ref{sec5}, we are able to compute solvability
conditions around this critical value of $\epsilon$ that lead to time dependent
amplitude equation. These amplitude equations describe a saddle--node
bifurcation. In conclusion, in section \ref{sec6}, we try to match this phase
approach, which formally cannot  describe vortices in NLS, with the release of
vortices observed in  simulation using the full nonlinear Schr\"{o}dinger
equation.

\section{Formulation of the problem}  
\label{sec2}

The Gross--Pitaevski\v{\i} equation reads, in a dimensionless form:

\begin{equation}
i \partial_t \psi({\bf x},t)  =  - \frac{1}{2}\nabla^2\psi +
\psi({\bf x},t) |\psi({\bf x},t)|^2 .
\label{nls}
\end{equation} 
This is a partial differential equation for a complex wave function $\psi({\bf
x},t)$. This equation is among other properties conservative and Hamiltonian. The
ground state in a periodic box is the homogeneous solution: $\psi_0 =
\sqrt{\rho_0} e^{-i \rho_0 t}$. Long wavelength and low amplitude perturbations
propagate with the sound  speed $c_s = \sqrt{\rho_0}$, on the other hand; 
$\xi_0 \sim \frac{1}{\sqrt{\rho_0}} = \frac{1}{c_s}$ is the only characteristic
microscopic length contained in this equation.

Writing $\psi = \rho^{1/2} e^{i\phi}$, we obtain two ``real'' hydrodynamical
fields, $\rho$ and $\phi$ representing respectively the particle fluid density
and the velocity potential:

\begin{eqnarray}
\partial_t \rho & = &
 -\nabla\cdot(\rho\nabla\phi); \label{cont} \\
\partial_t \phi & = &
-\frac{1}{2} (\nabla\phi)^2 +
 \frac{1}{2\rho^{1/2}} \nabla^2\rho^{1/2} - \rho . 
\label{bern}
\end{eqnarray}

The first one is the density mass conservation equation, identifying $\nabla\phi$ by
the local velocity $\bf v$. In the second equation, the term 
$ \frac{1}{2\rho^{1/2}} \nabla^2\rho^{1/2} $, often called quantum pressure is  
negligible for large scale flows, that is for flows with a space scale much larger than 
the intrinsic microscopic length, $\xi_0$. When 
this quantum pressure is neglected, the equation for $\phi$ is the equivalent of 
Bernoulli's equation for a compressible fluid, with an equation of state for the pressure
($p$): $p = \frac{1}{2}\rho^2$. For a stationary flow, $\partial_t \phi$ 
is a constant, defined by the conditions at infinity. Therefore, the
mass density $\rho$ can be computed everywhere as a  
function of $v$, the modulus of the velocity from this Bernoulli equation.

>From equations (\ref{cont}) and (\ref{bern}), we obtain the equations 
for the stationary flow around a disk
(of radius $R$, much bigger than any intrinsic length scale $\xi_0$) 
with a velocity at infinity $v_\infty$:

\begin{eqnarray}
\nabla\cdot\left(\rho(|\nabla\phi|) \nabla\phi \right) &=&0 ; \label{cont1} \\
\rho(|\nabla\phi|)&=&\rho_0+\frac{1}{2}(v_\infty^2-|\nabla\phi|^2)
\label{etat}\\
\hat n \cdot \nabla\phi &=&0 \quad  {\rm on \ the \ disk} \label{bord1}\\
\phi &=& v_\infty  x \quad  {\rm at \ infinity },
\end{eqnarray}
here $\hat n$ is normal to the disk perimeter.

Rewriting the equation (\ref{cont1}) in the frame reference defined by the pole
of the disk as the origin, one gets:

\begin{equation}
\partial_v (\rho(v)v) \partial_{xx}\phi + \rho(v) \partial_{yy}\phi = 0,
\label{phase}
\end{equation}
where $x$ is the local coordinate tangent to the main flow, and $y$ the
orthogonal one. At low velocities this second order partial differential equation
is elliptic. 
It is also possible to show, via an hodograph transformation, that the maximum local
speed for a flow around an obstacle, occurs on the boundary of the obstacle. This is in
some sense a nonlinear (but still elliptic) generalization of the min--max
theorem by Riemann and Liouville.
Equation (\ref{cont}) becomes hyperbolic beyond a critical velocity. This happens when
$\partial_v (\rho(v)v) $ vanishes\footnote{This equation becomes
hyperbolic also when $\rho(v) $ vanishes, however, this happens for a larger
value of the speed with the present equation of state (relation between
pressure and density).}, that is when the mass current takes its largest
possible value for some condition at infinite.

The above criteria ($\partial_v (\rho(v)v) =0$) gives, for the present model, a
critical velocity $v_c$ such that:  $v_c^2=\frac{2}{3}\rho_0+\frac{1}{3}
v_\infty^2$. When $v_\infty$ increases the property of ellipticity of equation
(\ref{cont}) is broken first at the  pole of the disk, leading to the nucleation
of two vortices, one at each pole.  As time goes on these vortices are convected
downstream by the mean flow. These vortices, once released, 
induce as well a counterflow because of the circulation condition
and this counterflow reduces the velocity on the surface of the disk. This
brings back the local velocity at the pole of the disk 
below the critical speed, and restores there the ellipticity of 
the equation for the velocity potential. But the vortex is pulled farther and 
farther downstream, and the counter streaming effect diminishes, 
till the velocity at the pole reaches eventually the critical value, the conditions at 
infinity being kept constant; then
new vortices are emitted, {\it etc.} This describes a more or less periodic
release of vortices from the obstacle \cite{fpr}. We shall study in this article
the process of nucleation, namely the way in which a  vortex is emitted from the
boundary when the local velocity  changes slowly from below critical to above
critical speed. For this we developpe a phase--dynamics approach
for the longwave asymptotics (distances larger than $\xi_0$). 

\section{The Euler--Tricomi equation near the transonic region.}
\label{sec2.5}

 Suppose that
the local velocity at the pole of the disk, $v_0$, is close from $v_c$; taking this point 
the origin of the axis, the $x$-axis being tangential to the disk, and the $y$-axis 
perpendicular. One writes the phase near the pole as $\phi=v_0 x + \frac{v_c}{3}
\chi$,
$\chi$ being small; this gives for equation (\ref{cont1}), together with
(\ref{etat}) (after an elementary redefinition of variables):

\begin{equation}
-(\epsilon + \partial_x \chi) \partial_{xx} \chi+  \partial_{yy}
\chi=0
\label{pha1}
\end{equation}
with $\epsilon\sim(v_0-v_c)/v_c$. The boundary condition (\ref{bord1}) becomes:
$$
(x/R,1)\cdot\left(v_0+\frac{v_c}{3}\partial_x\chi,\frac{v_c}{3}\partial_y\chi\right)=0 
\quad {\rm at} \quad y=-\frac{x^2}{2R};
$$
where $(a,b)\cdot(a',b')=aa'+bb'$.
Neglecting $\partial_x\chi$ in the boundary conditions gives:
\begin{equation} \partial_y\chi=-M\frac{x}{R} \quad {\rm at} \quad
y=-\frac{x^2}{2R}.
\label{cond.bor}
\end{equation}
With $M=\frac{3v_0}{v_c}$, a constant proportional to the actual Mach number. Let
us also notice that the boundary condition arises on $y=-\frac{x^2}{2R}$, which
is a parabolic  approximation of the disk near the pole.

Equation (\ref{pha1}) derives from a variational
principle\footnote{This is a direct consequence from equation (\ref{cont1})
which is the
extremum of $E=\frac{1}{2}\int dx dy\rho(v)^2 = \frac{1}{2}\int dx dy
\left[\rho_0+\frac{1}{2}(v_\infty^2-(\nabla\phi)^2)\right]^2$.} 
with an energy
\begin{equation}
 E = \int dx dy\left[ - \frac{1}{6} (\epsilon + \partial_x \chi) ^3 +
\frac{1}{2}(\partial_y \chi)^2 \right].
\label{var}
\end{equation}

The solution of (\ref{pha1}) satisfying the equation and the boundary condition is
$$\chi_0 = -M \frac{xy}{R},$$ 
however, this one is not sufficient to determine the complete flow in particular
the transition to supersonic flow, it is necessary to go up to next order.
Writing
$\chi= \chi_0 + \varphi$, one finds:

\begin{equation}
-\left( \epsilon - M \frac{y}{R}\right)\partial_{xx}\varphi+
 \partial_{yy}\varphi=0
\label{phase1}
\end{equation}
with
\begin{equation}
\partial_y\varphi=-M\frac{x^3}{R^3}; \quad {\rm at} \quad  y=0.
\label{bord}
\end{equation}
The resulting equation (\ref{phase1}) for $\varphi$ is
Euler--Tricomi\footnote{Note that here the variables are in the physical space
and not the hodograph variables as in
\cite{landau}. With this procedure we have considerate directly the boundary conditions,
something difficult to work with in the hodograph plane.}
with the boundary condition (\ref{bord}).

The Euler--Tricomi equation may be interpreted as follows:
$-(\epsilon-M\frac{y}{R})$ represents a generic tangential velocity profile
of an ideal flow near a body, since the local main speed diminishes as $y$
increases, that is as one moves far--away from the obstacle. The Mach number is
exactly one at $y=\epsilon\frac{R}{M}$.
In this equation we have neglected the nonlinear term
$\partial_x\varphi\partial_{xx}\varphi$ besides 
$y\partial_{xx}\varphi$ in (\ref{phase1});
on the other hand the border of the obstacle has been taken at $y=0$.
(The curvature of the obstacle brings a correction to the boundary condition
for the dominant order solution, that transforms itself to an extra term in
the partial differential equation for the perturbation with a flat boundary 
$\varphi$.) 
This assumption is consistent to the following scaling in
$\epsilon$ for the coordinates 
\begin{equation}
x \sim \frac{R\epsilon^{3/2}}{M} \quad , \quad  y
\sim \frac{R\epsilon}{M} \quad {\rm and} \quad 
\varphi \sim
\frac{R\epsilon^{11/2}}{M^3}.
\label{scaling}
\end{equation}

Once again, there exists a particular solution of equation
(\ref{phase1}) satisfying the boundary conditions,
so that we will have to focus on the homogeneous solution of its equation.
It happens that, $\phi_0$, defined as:
$$ \phi_0=-M\frac{x^3y}{R^3}-\epsilon M \frac{xy^3}{R^3}+M^2 \frac{xy^4}{2R^4}$$
satisfies equation (\ref{phase1}) with the (\ref{bord}) boundary conditions.

At this point, one can iterate the linear Euler--Tricomi equation by the same
procedure, that is by considering how $\phi_0$ modifies the true boundary
condition and then we compute the next term $\phi_1$, and so on.
It is then possible to find a polynomial expansion satisfying the full boundary
condition. However, the convergence of this expansion is
not obvious. This is not an objection in principle because we are
looking only local solutions, however we have in mind an outer
asymptotics which will mtach with a vortex kind solution (see section
\ref{sec6}.

\section{Special solutions of the Euler--Tricomi equation}
\label{sec3}

Let us first make an ``aparte'' by looking at some special solutions of 
equation (\ref{phase1}) that might give some idea of what arises when
crossing the critical velocity, that is at the transonic transition.

The roots $z(x,y)$ of the cubic polynomial (as well as any linear
combination of these three roots):
\begin{eqnarray}
 z^3+3\left(\frac{M}{R}\right)^{1/3}\left(y-\epsilon\frac{R}{M}\right) z +3x
\label{polyn}
\end{eqnarray}
are exact solutions of equation (\ref{phase1}).
Note that for $y>\epsilon\frac{R}{M}$,
the cubic polynomial has only one real root for all values
of $x$, whereas, for $y<\epsilon\frac{R}{M}$ there are three real roots
inside a semi--cubic parabola defined by
 $|x| \leq 2/3 (\epsilon-y)^{3/2}$.

\begin{figure}
\centerline{\epsfxsize=8truecm \epsfbox{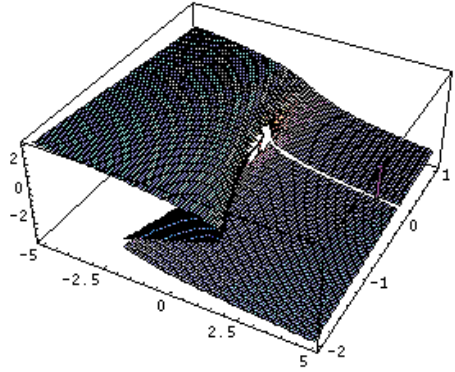} }
\caption{\protect\small Shape of $z(x,y)$, for $\epsilon=0$ around the
origin.
\label{multi}}
\end{figure}

This multivaluation of the real roots of the cubic equation (\ref{polyn})
means that it is not possible, generally, to follow continuously a root of
(\ref{polyn}) along a closed path around the origin (see figure (\ref{multi})).
More precisely, such solution will admit a discontinuity in the region inside
the semi--cubic.
Unless one can regularize the discontinuity which arise along the region
multivalued (in the same sense that we have to regularize the over
turning of waves dynamics, solutions of nonlinear and non dispersive wave
equations), there is no hope of having a stationary solution of our problem
except if the discontinuity gap is
$2\pi$. In this case (jump of $2\pi$ of the phase), we will see that even if
both the quantum pressure and the limited transonic region have a tendency to
restore the smoothness of the solution, there is a critical velocity above which
there is no more  possibility of having stationary solutions. The way we will
treat the  equation will hide the multivalued property that we pointed out
because  we will look on regular stationary solutions (which exist as well as no 
discontinuity appears), so that the discontinuities will be solved {\it via}
the general time dependent nonlinear problem.

A general solution arises whenever one consider a continuous superposition in
the neutral translation mode of the above solution (\ref{polyn}):
$$\phi(x,y) = \int a_\xi z(y, x-\xi) d\xi$$ 
equivalent after a change of variable to the integral expresion in \S 118 of
\cite{landau}. This general expression is valid only for
$y>\epsilon\frac{R}{M}$ (as well as the roots of the polynomial (\ref{polyn}))
where such a change of variable remains well defined.

Before finish this section, let us mention, that another family of solution
appears when one takes the Fourier transform of the $x$ variable. Then one has
that $$\phi_\nu(x,y) =\Phi(y)e^{i\nu x},$$
with $\Phi(y)$ is the Airy function, the detailed analysis of this kind ok
solution is elaborated below in a more general way.

\section{Regularization of the shock solutions of the Euler--Tricomi
equation by the quantum pressure.}
\label{sec4}

As soon as the solutions of the Euler--Tricomi equation appears to be sharper
and sharper that the quantum pressure term in the Bernoulli equation
(\ref{bern}) is no longuer negligible because it involves higher order
derivatives. From the full Bernoulli equation one gets the value of $\rho$ by an
implicit relation (we shall consider here also the role of non stationary
dependence of phase and density in order to capture the full short wavelength
dynamics):
$$ \rho=\rho_0+\frac{1}{2} (v_\infty^2-({\bf \nabla}\phi)^2)-\partial_t \phi
+\frac{1}{4\rho} \left(\Delta \rho -\frac{ ({\bf \nabla}\rho)^2}{2 \rho}\right),
$$ reminding that $\phi= v_0 \cdot x+ \frac{v_c}{3}(\chi_0+\phi_0+\varphi) $
$\phi$ and $\varphi$ being function of both time and position whereas 
$v_0$, $\chi_0$ and $\phi_0$ are independant of time. One can then estimate 
the value of the quantum pressure at the first order of perturbation, taking
 $\rho=\rho_0+\frac{1}{2} (v_\infty^2-({\bf
\nabla}\phi)^2)-\frac{v_c}{3}\partial_t \varphi$ (it gives then  the first
non--zero contribution of the quantum pressure);  coupling this with the (now)
non--stationnary mass conservation equation and  restoring the first nonlinear
terms as well as the constant terms gives for the phase equation:

\begin{equation}
-\left(\epsilon-M\frac{y}{R}\right)\partial_{xx} \varphi+\partial_{yy} \varphi
-\xi_0^2 
\partial_{x^4}\varphi= 
\frac{1}{v_c^2} \partial_{tt} \varphi+ \frac{M}{v_c} \partial_{tx}\varphi
+\partial_x \varphi \partial_{xx} \varphi
+\partial_x \phi_0
\partial_{xx} \varphi+\partial_{xx} \phi_0\partial_x \varphi
+\partial_x \phi_0 \partial_{xx} \phi_0.
\label{finale}
\end{equation}

We have kept in this equation only the most important term of each contribution.
The quantum pressure term, $\xi_0^2 \partial_{x^4}\varphi$ should be 
multiplied by a number that we have taken 
to one by simplicity.

The way this equation (\ref{finale}) is written is dictated by our method 
of resolution: the left hand side will be in fact treated as the main 
equation, linear and homogeneous, while the right hand side corresponding to
perturbations which will be incorporated terms by terms. As said before the
boundary conditions are taken homogeneous:
$$
\partial_y \varphi = 0 \quad {\rm at} \quad y=0.
$$

Let us study the regularisation of the solution above the threshold,
first by adding the quantum pressure then by considering cross terms 
involving $\varphi$ and $\phi_0$; at these points the analysis remains linear 
so that we will just be able to look to homogeneous solutions of
our problem, without solving for the amplitude; finally we will focus on 
the global nonlinear problem which allows to calculate amplitude equations and
leads to the study of the time--dependent evolution.

First we look at the left hand side of equation (\ref{finale}), taking the right
hand side as zero. The boundary conditions are $\partial_y \varphi=0$
at $y=0$, so that one can expand the solution as a linear superposition 
of functions. The equation can be solved as the Euler--Tricomi one,
using the Airy function.

\begin{figure}
\centerline{ \epsfxsize=10truecm \epsfbox{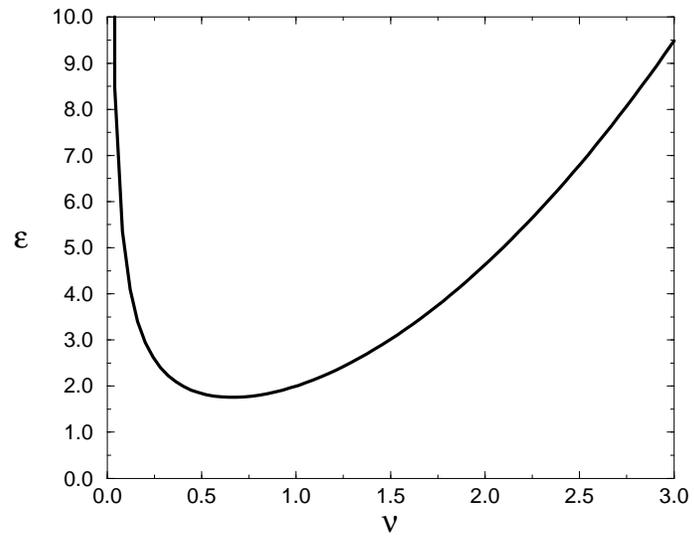} }
\caption{\protect\small Relation between
$\epsilon$ and $\nu$ the wave number for $R/\xi_0 \simeq 1$.
\label{quant}}
\end{figure}

We seek solutions of the form $\varphi_\nu = e^{\pm i\nu x} \zeta(y) $,
$\zeta(y)$ satisfying the Airy equation:
$$ \zeta''+ \nu^2\left(\epsilon-\xi_0^2\nu^2-M\frac{y}{R}\right)\zeta =0$$
which non-divergent solution is known to be the Airy function $\Phi(\cdot)$;
 therefore the solution reads:
$$ \varphi_\nu=A \cdot e^{i\nu x}
\Phi\left[\left(\frac{\nu^2M}{R}\right)^\frac{1}{3}
\left(y-\frac{R}{M}(\epsilon-\nu^2\xi_0^2)\right)\right].$$
being $A$ a complex amplitude fixed by nonlinearities at next order.
As the function $\Phi(s)$ does not possess extrema for $s>0$ the boundary
condition might be  satisfied for $\epsilon>0$ only, otherwise  $A=0$. Let
$s_n$ be the $n^{th}$ zero of
$\Phi'(s)$ then
the only possible values for the wavenumber $\nu$ are such that they satisfy a
``quantization condition'' for a given $\epsilon$:
\begin{eqnarray}
\epsilon=\left(\frac{M \xi_0}{R}\right)^\frac{2}{3}(-s_n)
\left(\xi_0 \nu\right)^\frac{2}{3}+(\xi_0 \nu)^2
\label{quant2}
\end{eqnarray}
which have been represented on figure \ref{quant}. So, if $\epsilon$ is less
than a critical value, one can observe that, because of the quantum pressure,
the homogeneous Euler--Tricomi equation has just the null function as solution
($A=0$). This critical value (the minimum  of the curve plotted in figure
(\ref{quant}) is easy to evaluate:
$$\epsilon_c=4 \times \left(-\frac{ s_1}{3}\right)^\frac{3}{4}\times
\sqrt{\frac{M\xi_0}{R }}=4 \xi_0^2 \nu_c^2 \quad
{\rm with} \quad \nu_c=\left(-\frac{ s_1}{3}\right)^\frac{3}{8} \times
\left(\frac{M }{R\xi_0^3}\right)^\frac{1}{4}$$
 the critical wave number.
Generically, this means that for $\epsilon < \epsilon_c$ the 
stationary solution ($\phi_0$) might describe the dynamics at the order of the 
Taylor expansion whereas for $\epsilon \geq \epsilon_c$ the amplitude of a
solution for $\nu=\nu_c$ can expand. At this point, the amplitude cannot be
known and has to be found as the result of the nonlinear and time dependent
analysis. It follows now a general scheme, which consists of
the evaluation of the amplitude and the corrections of the general solution
by writing a solvability condition for their existence.

We seek a solution of the form: 
$$\phi= A(x) \cdot e^{\pm i\nu_c x}\Phi\left[ \left(\frac{\nu_c^2
M}{R}\right)^\frac{1}{3}
\left(y-\frac{3R\epsilon_c}{4M}\right)\right]+\varphi_1(y)e^{\pm i\nu_c x}.
$$
Now $A(x)$ is a slightly varying amplitude on the horizontal variable
$x$ (the time dependance will be taken account in the next section)
$\varphi_1$ being a small correction to $\varphi$ depending on $y$ only 
(the $x$ dependance coming from the main term $e^{i\nu_c x}$).
This first correction of our equation (\ref{finale}) will takes in 
account the cross term between $\varphi$ and $\phi_0$; the slow behavior of 
$A(x)$ on $x$ requires:
$$\left|\frac{\partial_x A(x)}{A(x)}\right| \ll \nu_c.$$

In addition let us define the $\zeta(y)$, the
Airy function that we use: $\zeta(y)=\Phi( (\frac{\nu_c^2 M}{R})^\frac{1}{3} 
(y- \frac{3R\epsilon_c}{4M}))$.
The cross contribution to the nonlinear
term $\phi_x \phi_{xx}$ which we have neglected until now, give two terms linear
in $\varphi$ in equation (\ref{finale}): $\partial_x \phi_0 \partial_{xx}
\varphi$ and $\partial_{xx} \phi_0 \partial_x \varphi$.

Therefore we obtain for the phase the following equation:
\begin{eqnarray}
{\cal L}_0 \varphi_1&=[\nu_c^2(-(\epsilon-\epsilon_c)-\partial_x\phi_0)A(x)
+i \nu_c \partial_{xx} \phi_0 A(x)+&
2i\nu_c(\frac{\epsilon_c}{2}-M\frac{y}{R}+\partial_x \phi_0)A'(x) \nonumber \\
&+\partial_{xx} \phi_0 A'(x)
-(\frac{\epsilon_c}{2}+M\frac{y}{R}-\partial_x \phi_0)A''(x)]\zeta(y) &
\label{eqsol}
\end{eqnarray}
where ${\cal L}_0$ is the linear operator acting on the one variable function 
space:
$$ {\cal L}_0 = \partial_{yy}+\nu_c^2\left(3\xi_0^2 \nu_c^2 -
M\frac{y}{R}\right).
$$  
Notice that $\zeta(y)$ is in the kernel of ${\cal L}_0$.
The solvability condition says that the right hand side of (\ref{eqsol}) 
is orthogonal to members of the kernel of ${\cal L}_0$ joint operator. With the
scalar product
$\left<f,g\right>=\int_0^\infty f(y) g(y) dy$ the linear operator
${\cal L}_0$ is self--adjoint, therefore 
$\zeta(y)$ belongs to its kernel. The solvability condition of
(\ref{eqsol}) gives an equation for the slowly varying amplitude (keeping the
first order in $A$ only)
$A(x)$:
$$ A''(x)+ \left( \frac{\epsilon-\epsilon_c}{4\xi_0^2}-\frac{3\nu_c^2
x^2}{2R^2}\right) A(x)=0 .$$
One recognize the equation of the quantum harmonic oscillator; 
$ \frac{\epsilon-\epsilon_c}{4\xi_0^2} $ being the equivalent of the energy.
It has non zero solution exists if $\epsilon> \epsilon_c'$ only. The lowest value
of epsilon with non--zero solution satisfies:
$$ \epsilon_c'-\epsilon_c=\sqrt{6\epsilon_c}\frac{\xi_0}{R} \ll \epsilon_c.$$
Note that this correction is small with respect to the first one.
The corresponding solution at $\epsilon = \epsilon_c'$
reads 

$$A(x)=A e^{-\frac{(\epsilon_c'-\epsilon_c)x^2}{8\xi_0^2}}=e^{-\frac{x^2}{2
l^2}}.$$
being $l$ the characteristic length of $A(x)$, $l=2 \sqrt{\xi_0 R}
/(6\epsilon_c)^\frac{1}{4}$ (note that this agrees with the condition of
validity of this WKB approach $\nu_c  l\gg 1 $).
As for the former treatment, a non zero amplitude $A$ can arise only if 
$\epsilon>\epsilon_c'$; this amplitude and its dynamics will be obtained by the
time dependent nonlinear system for $\epsilon \sim \epsilon_c'$.

\section{Amplitude equation for the saddle--node bifurcation}
\label{sec5}

In this section we shall look at the general spatio--temporal dependence of the
solution of (\ref{finale}) for 
$\epsilon \sim \epsilon_c'$; writing now the {\it Ansatz} for the phase
$\varphi$ as ($A(t) = \alpha(t)+i\beta(t)$):

$$\varphi=(\alpha(t)\sin(\nu_c x)+ \beta(t) \cos(\nu_c x))e^{-\frac{x^2}{2
l^2}}\zeta(y)
+\varphi_1(y) \cos(\nu_cx+\Omega)+\varphi_2(x) \zeta(y),$$
being $\varphi_1(y)$ and $\varphi_2(x)$ small corrections to $\varphi$.
Imposing a solvability condition (one first along $y$, then along $x$), one
gets the following dynamical system:

\begin{eqnarray}
\frac{1}{v_c^2} \alpha''(t)-M \frac{\nu_c}{v_c}\beta'(t)&=&\nu_c^2(\epsilon-\epsilon_c') \alpha(t) + \delta
e^{-\frac{\nu_c^2l^2}{2}}+\nu_c^3(\alpha^2(t)-\beta^2(t))e^{-\frac{\nu_c^2l^2}{6}} \nonumber \\
\frac{1}{v_c^2} \beta''(t)+M \frac{\nu_c}{v_c}\alpha'(t)&=&\nu_c^2(\epsilon-\epsilon_c') \beta(t) + 2\nu_c^3 \alpha(t)\beta(t)
e^{-\frac{\nu_c^2l^2}{6}} \label{amplitude}
\end{eqnarray}
with $\delta \sim \xi_0 /R^2 $. For convenience, and because it does not 
change the meaning of the dynamics, we have taken the ratio between the 
different constant coming from solvability integration equal to one. The 
transcendental terms (in $e^{-\nu_c^2l^2}$) are due to the interaction between 
the nonlinear term and the constant term $\partial_x \phi_0 \partial_{xx} 
\phi_0$ with the modes $\sin(\nu_cx)e^{-\frac{x^2}{2
l^2}}$ and 
$\cos(\nu_cx)e^{-\frac{x^2}{2
l^2}}$.

The system (\ref{amplitude}) could be written in a single complex equation for
the complex amplitude (after an appropiate change of variable and rescaling):
\begin{equation}
 Z_{tt}+i\omega Z_{t}=\mu + Z^2.
\label{complex}
\end{equation}
Here $\mu =  - 
(\epsilon-\epsilon_c')^2e^{\frac{ \nu_c^2 l^2}{3}}/4\nu_c^2+ \delta 
e^{-\frac{ \nu_c^2 l^2}{3}}/\nu_c^3$.

The stationary solutions are $Z_\pm = \pm\sqrt{-\mu}$
that is for negatives values of $\mu$ one preserves an odd symmetry of $
\varphi(x,y) $ along the $x-$axis, this symmetry is broken as soon as $\mu$
changes sign and the further evolution is more complex.

In terms of the physical parameters the region $\mu <0$ is for
$(\epsilon-\epsilon_c')^2> \frac{4 \delta}{\nu_c}
e^{-\frac{2 \nu_c^2 l^2}{3}}$, leading to a new critical velocity 
$\epsilon_t$, defined as:
$$ \epsilon_t=\epsilon_c'-2
\sqrt{\frac{\delta}{\nu_c}}e^{-\frac{\nu_c^2l^2}{3}}.
$$
Notice that for small aspect ratio, the main correction for the critical
velocity is in $\sqrt{\frac{\xi_0}{R}}$ and is determined by $\epsilon_c$
whereas $\epsilon_c'$ and $\epsilon_t$ induce just exponentially small 
corrections.

\begin{figure}
\centerline{ \epsfxsize=10truecm \epsfbox{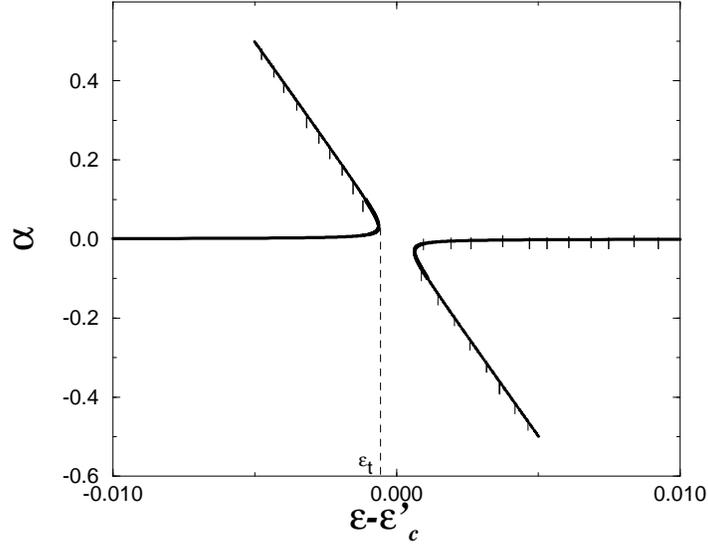} }
\caption{\protect\small Stationary amplitude of the phase equation for
$\frac{\xi_0}{R}=0.1$. The part for $\epsilon>\epsilon_c'$ is not valid in
our approximations whereas we have dashed the unstable solutions.
\label{diag}}
\end{figure}

\begin{figure}
\centerline{ \epsfxsize=10truecm \epsfbox{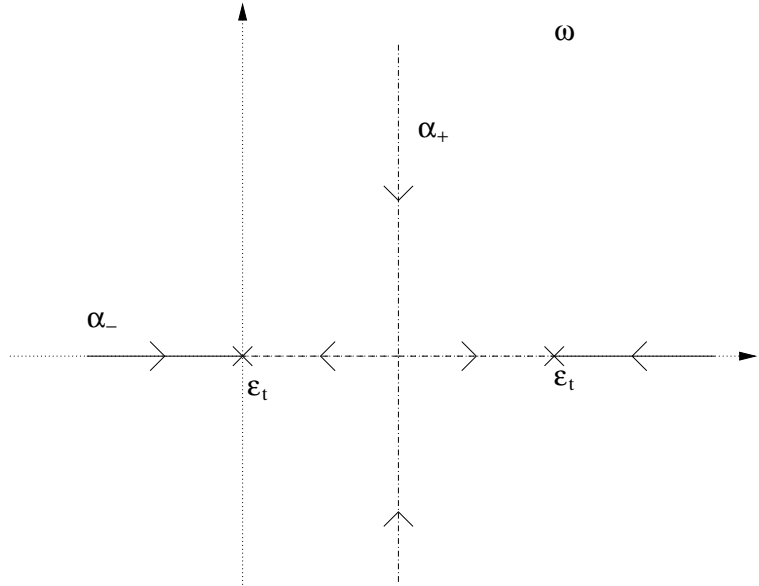} }
\caption{\protect\small Stability diagramm of the stationary solutions
$\alpha_\pm$ in the $\omega$ complex plane (the stability is studied via the
expression $w(t)= a \cdot e^{i\omega t}$, $a$ being an amplitude.
The small dashed line corresponds to the real and imaginary axes whereas the
large dashed line indicates the stability evolution of $\alpha_+$ as $\epsilon$
increases untill $\epsilon_t$ where the stationnarity disappears as the collapse
of $\alpha_+$ and $\alpha_-$. The thick line represents the same evolution for
$\alpha_-$.
\label{dess}}
\end{figure}

Figure (\ref{diag}) shows the real roots of $Z_\pm$ as a function of $\mu$ or
better as a function of
$\epsilon-
\epsilon_t$. One can identify two branches: one for low velocities ($\epsilon <
\epsilon_t$); the other one for high velocities. As we have conducted our 
calculations, starting from stationary solution for low velocity and then 
adding the nonlinear dynamics step by step, this second branch for high velocity
is irrelevant in our problem.
As shown on figure (\ref{diag}), for $\epsilon<\epsilon_t$ there are two
roots of the stationary equation, roots that greather for $\epsilon=
\epsilon_t$, whereas there is no more roots for $\epsilon$ greater than 
$\epsilon_t$, giving rise to a saddle--node bifurcation.
The linear stability analysis of the stationary solutions $Z_\pm$ gives the
following dynamic ($Z(t)=Z_\pm+z(t)$, and $|z(t)|\ll |Z_\pm|$):
$$ z_{tt}+i\omega z_{t}=\pm 2\sqrt{-\mu}z.
$$
Therefore, one notes that for:
$$ \epsilon<\epsilon_c'-\sqrt{\frac{M^4}{16}+4 \frac{\delta}{\nu_c}
e^{-\frac{2}{3} \nu_c^2 l^2}} $$
one branch is stable ($Z_-$, representing by $A_-$ on figure (\ref{diag})) 
and the other is always unstable as in ordinary saddle--node bifurcations. 
However, because of to the oscillatory term $i\omega z_{t}$, when approaching
the bifurcation, both solutions are stable. This happens in the range of values
of $\epsilon$ such a that
$$ \epsilon_c'-\sqrt{\frac{M^4}{16}+4 \frac{\delta}{\nu_c}
e^{-\frac{2}{3} \nu_c^2 l^2}}<\epsilon<\epsilon_t $$
The resulting phase diagram of the bifurcation has been plotted on figure 
(\ref{dess}).
We argue that this saddle-node bifurcation gives a consistant scenario of
the vortex nucleation seen in figure (\ref{photo}). They appear as a
consequence of the disappearence of stationnary solutions on the phase 
dynamic approch.
 The vortex
diminishes the local velocity so that the flow comes back to a description
valid for $\epsilon<\epsilon_t$ and when the vortex is far enough, because
of the advection due to the mean velocity, $\epsilon$ pass again through the
transition and we got a periodic vortex-nucleation process. The saddle--node 
bifurcation appears in our case to be richer because close to the transition 
both stationnary solutions are stable. This result is in
a good agreement with numerical solution by Huepe and Brachet
\cite{huepe}.

\section{Breakdown of the phase description, the appearance of vortex
motion, conclusions and further miscellaneous}
\label{sec6}

We have in the former section reviewed the phase description, that explains the
disappearence of any stationary solution for the flow problem. This is the
first step toward the nucleation of a vortex. Briefly, one can retain that this
periodic behavior appears as a saddle--node bifurcation  where two branches of
stationnary solutions collapse giving rise for larger  velocities to non
stationnary solution at all. The numerous critical parameters we have mentionned
might be simplified if one see that as $R \gg 
\xi_0$, we have $\epsilon_t \sim \epsilon_c' \sim \epsilon_c$. This allows to
claim that the critical velocity for vortex nucleation $v_v$ reads at first
order correction of the critical speed:
$$ v_v-v_c \propto \sqrt{\frac{\xi_0}{R}} +{\cal
O}\left(e^{-\nu_c^2l^2}\right).$$ 
where $v_c$ is the Landau critical velocity.

Finally, it remains to match the vortex
nucleation as a process being part of the same evolution of the dynamical
saddle--node bifurcation. For simplicity we will consider the amplitude
equation (\ref{complex}) where we have dropped the first order time derivative,
a term much smaller than the other ones as one approaches the time when
nucleation occurs (see the scaling below):
$$ \ddot Z(t)=Z^2(t)+\mu $$
where $\mu$ is related to $\epsilon-\epsilon_t$. When $\mu$ is negative,
there is two real stationnary solutions ($Z=-\sqrt{-\mu}$ the stable, 
$z=\sqrt{-\mu}$ the unstable) corresponding to $\alpha_\pm$. 
The saddle-node bifurcation is crossed when $\mu$ becomes positive. This
can be studied by taking $\mu=t$ (by rescaling, no multiplying factor is
needed) and then we obtain the first Painlev\'e transcendent\footnote{One
may note that taking $\mu=t$ means to take $\epsilon = \dot \epsilon t$ in
equation (\ref{finale}). This gives
$$-\left(\dot\epsilon t -M\frac{y}{R}\right)\partial_{xx} \varphi+\partial_{yy}
\varphi -
\frac{1}{v_c^2} \partial_{tt} \varphi - \frac{M}{v_c} \partial_{tx}\varphi=0,$$
which, after a change of variables of the form: $\eta = \dot\epsilon t -My/R$ and
$\zeta = Mv_0t/R + \dot\epsilon y/v_0$, and neglecting the linear
derivative in time as in (\ref{pain}), leads to:
$$ -\eta\partial_{xx} \varphi+ \left(M^2/R^2-\dot\epsilon^2/v_0^2\right)
(\partial_{\eta\eta}\varphi - \partial_{\zeta\zeta} \varphi)= 0.$$
A kind of Euler--Tricomi equation, interesting by itself. One note that,
if the rate of acceleration at infinity $\dot\epsilon$ is larger
than $Mv_0/R$, the nucleation process is caused by a dynamical instability not by
a sonic transition as is described in this article. }\cite{Ince}:
\begin{equation}
\ddot Z=Z^2+t
\label{pain}
\end{equation}

A convenient change of variable, for $t<0$ is:
$$z=\sqrt{-t}W(T);\quad \quad T=\frac{4}{5}(- t)^{5/4} $$
and gives the following equation:
$$ \ddot W+\frac{\dot W}{T}-\frac{4}{25}\frac{W}{T^2}=W^2-1. $$
It typically gives the two stationnary solutions $W=\pm1$ for large $T$ and
it can be solved in terms of Weierstrass function for large $T$ \cite{Ince}. 
Equation (\ref{pain}) is known for giving finite--time singularities which
shows that the slowly varying approximation of the phase dynamics breaks at
some point. In addition, one can notice that
$Z$ is related to the velocity of the  fluid in the moving frame (of velocity
$v_0$). Then, one expect to relate the amplitude $Z$ to a order parameter which
parametrizes a continuous family of solutions of the full nonlinear Sch\"odinger
equation (\ref{nls}). Jones and Roberts 
\cite{jones} found the kind of solution that we are interested, in
two and three spatial dimensions, consisting in axisymmetric solitary
structures. The solution proposed is

\begin{equation}
\phi= - U x + m\frac{2x(1- U^2) }{ x^2 +(1-U^2)y^2}.
\label{rob}
\end{equation}
$m$ being a constant and $U$ the free parameter characterizing the whole family
of solutions.

Whenever the parameter (a velocity) $U$ of these solutions is close
to the unity (the sound speed) one identifies the relative speed $U-1$ (via a
change of variable) with the true order parameter $\alpha \& \beta$ of the
saddle--node bifurcation. 

On the other hand as $U$ goes to zero one matches the solution (\ref{rob})
with a pair of moving vortices, one with negative topological charge located
in  $(x=0, y=a)$ while the other is an image one inside the disk $(x=0,
y\approx -a)$\footnote{Note that at first order the dependence on the radii $R$
of the disk is not relevant to this solution.}:
\begin{eqnarray}
\phi&=& - \frac{U x}{\sqrt{1- U^2} } + \arctan\left(\frac{ (y+a)\sqrt{1-
U^2}}{x} \right) -
 \arctan\left(\frac{ (y-a)\sqrt{1- U^2}}{x} \right) \nonumber\\
&\approx & - \frac{U x}{\sqrt{1- U^2} } + \frac{2xa\sqrt{1- U^2} }{ x^2
+(1-U^2)y^2}.
\label{rob1}
\end{eqnarray}
 
One relates very easily the distance $a$ in (\ref{rob1}) with the speed $U$ in
(\ref{rob}) by $U=\frac{1}{2a}$: the Hemholtz law of motion for point--like
vortices. As $a$ diminishes, a vortex pair appears. So in the dynamical context
$a$ satisfy formally the same equation than the quantity $z$ of (\ref{pain}) or
$\alpha$ of (\ref{amplitude}), when the time dependant solution evolves (after
crossing the saddle--node bifurcation), a vortex appears as
$\alpha$ increases, the second vortex of the pair being formally inside the disk
in other to preserve the boundary condition.

Finally, it is of a general interest to note that the shallow water equations
have exactly the same shape that the set (\ref{cont},\ref{bern}) in the case of
potential fluid motion (see \cite{landau}). In fact the shallow water equations
are
\begin{eqnarray}
\partial_t h & = &
 -\nabla\cdot(h {\bf v}); \nonumber \\
\partial_t {\bf v} + {\bf v}\cdot\nabla {\bf v} & = & - g\nabla h . 
\nonumber
\end{eqnarray}
After imposing a vortexless flow: ${\bf rot} \ {\bf v} =0$, {\it i.e.} ${\bf v}
=\nabla\phi$, and neglecting the quantum pressure term in (\ref{bern}), one
identifies the height of fluid
$h$ with the superfluid density $\rho$, and the potential fluid velocity with the
phase of the condensate wave function. It is possible to get a short scale term
in the shallow water equations by adding a capillary term, therefore 
it exist a complete
analogy with the analysis developed {\it in extenso} in this article. Perhaps a
transonic transition with the predicted scenario is observable in mercury where
the kinematic viscosity is very low. However, the final state will be
different because the deep significance of the phase of the wave function
$\psi$ does not extend the existence of quantized circulation vortices for the
case of surface waves, where as it is well known the circulation takes any
value.

\eject

\eject

\eject

\eject

\begin{figure}
\centerline{ \epsfxsize=8truecm \epsfbox{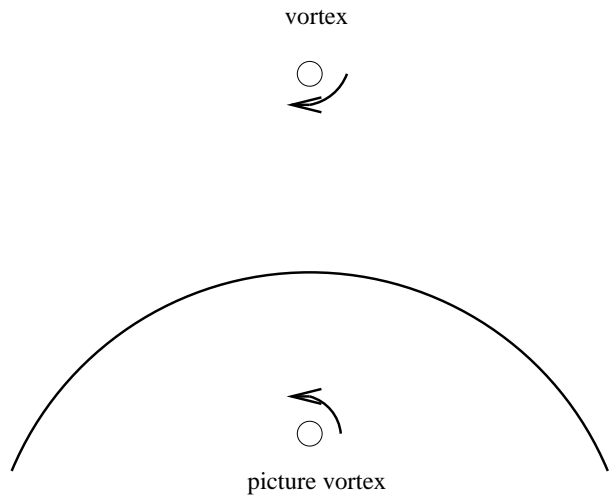} }
\caption{\protect\small Position of the vortices for the unstable stationnary
solution.
\label{vortex}}
\end{figure}

\end{document}